# The heteronomy of algorithms: Traditional knowledge and computational knowledge[i]

David M. Berry

1   If an active citizen should increasingly be a computationally enlightened one, replacing the autonomy of reason with the heteronomy of algorithms, then I want to argue in this chapter that we must begin teaching the principles of critiquing the computal through new notions of what we might call digital *Bildung*.[1] Indeed, if civil society itself is mediated by computational systems and media, the public use of reason must also be complemented by skills for negotiating and using these computal forms to articulate such critique.[2] This critical spirit of majority also serves to problematize the idea that the university is only useful for producing mandarins and workers, and highlights the continuing importance of critical thinking in the humanities and social sciences in a digital age.[3] Not only is there a need to raise the intellectual tone regarding computation and its related softwarization processes, but there is an urgent need to attend to the likely epistemic challenges from computation which, as presently constituted, tends towards justification through a philosophy of utility rather than through a philosophy of care for the territory of the intellect. Indeed, the mechanization of mind, long an aim of the computational sciences, is now at hand in a number of moments. Human thinking is too often conceptualized through an instrumentalist rationality that seeks to undermine the very possibility of rational critical thinking in an age when that very critical rationality is urgently needed. An example of which is current debates about Big Data and its capture, processing, storage, and feedback into our thinking and behavior in order to prescribe specific effects, such as the Quantified Self movement which uses such data to regulate individual behavior.[4]

2   Today there are rapid changes in social contexts that are made possible by the installation of code/software via computational devices, streams, clouds, or networks, what Mitcham calls a "new ecology of artifice."[5] The proliferation of computational contrivances that are computationally based has grown rapidly, and each year there is a large growth in the use of these computational devices and the data they collect. These devices, of course, are not static, nor are they mute, and their interconnections, communications, operation, effects, and usage are increasingly prescriptive on the everyday life world. But as opaque devices they are difficult to understand and analyze due to their staggering rate of change, thanks to the underlying hardware technologies, which are becoming ever smaller, more compact, more powerful and less power-hungry; and by the increase in complexity, power, range and intelligence of the software that powers them. Within the algorithms that power these devices are embedded classificatory schemes and ontologies that pre-structure the world that is presented. Indeed, this formatting and mediating capacity directly encodes cover concepts into the device.

3   Through the introduction of softwarized technical systems, it is sometimes claimed that we live in an information society.[6] Whilst numerous definitions exist, we now appreciate that all around us software is running on digital computers in an attempt to make our lives more comfortable, safer, faster, and convenient—although this may conversely mean we feel more stressed, depressed, or empty of meaning or purpose due to our new softwarized world. Indeed, it seems more accurate to state that we live in a *softwarized* society. From the entertainment systems we use to listen to music and watch television, to the engine systems that allow us to experience unprecedented fuel efficiency and even electric cars, to the computer modelling that manages the banking system and even the entire economy, software is doing the heavy lifting that makes the old industrial methods anachronistic. We therefore need to develop an approach to this field that uses concepts and methods drawn from philosophy, politics, history, anthropology, sociology, media studies, computer science, and the humanities more generally, to try to understand these issues—particularly the way in which software and data increasingly penetrate our everyday life and the pressures and fissures that are created. We must, in other words, move to undertake a critical interdisciplinary research program to understand the way in which these systems are created, instantiated, and normatively engendered in both specific and general contexts.[7]

4   In addition to the proliferation of computation and computational practices, we are starting to see changes in the way we understand knowledge, and therefore think about it. Computation is, in many cases, fundamentally changing the way in which knowledge is created, used, shared and understood, and in doing so changing the relationship between knowledge and freedom. Computation, and the data which it collects and produces, should encourage us to ask philosophical questions in a computational age and the relationship to the mode of production that acts as a condition of possibility for it. Indeed, following Foucault the "task of philosophy as a critical analysis of our world is something which is more and more important. Maybe the most certain of all philosophical problems is the problem of the present time, and of what we are, in this very moment… maybe to refuse what we are."[8] This call is something we need to respond to in relation to the contemporary reliance on computational forms of knowledge and practices and the co-constitution of new computational subjectivities. If critical approaches are to remain relevant in a computational age, then philosophy must work to critique and understand how the materiality of the modern world is normatively structured using computation and the attendant imaginaries made possible for the reproduction and transformation of society, economy, culture and consciousness.

5   However, these new digital technologies are not the sole driver of social and political change, rather, technology offers specific affordances within certain contexts which enable and disable certain forms of social and political interactions. Putting it another way, certain technologies within historical and social contexts serve to accelerate styles and practices of life, and marginalize others. But crucially they are also linked to associational structures of the specific network, organizational forms and processes used to achieve a certain "performance." To comprehend the digital we must, therefore, know it from the inside, we must know its formative processes. We can therefore think of technologies, and here I am thinking particularly of digital technologies, as being *embedded* in an important sense but also able to effect enframing processes through the element of agency that computational systems engender.

6   The speed and iteration of innovation in this area of technology might be incredibly fast and accelerating, but software *can* be materialized so that we may think critically about it. For example, it is important to recognize that software requires a platform upon which to run. New digital technologies form path dependencies that can become strengthened and naturalized as platforms, becoming self-reinforcing, creating a circle of technological leaps and accelerations. For example, new forms of knowledge platforms are built to structure our reading in particular ways, opening the possibility of distracted and fragmentary reading habits in contrast to deep reading, which may make it difficult to develop critical reflection or offer space for contemplation. Platforms can be either hardware or software-based, but they provide the conditions and environment which make it possible for the software to function correctly. The platform can offer a standpoint from which to study software and code, and hence the digital, but this approach is not sufficient without taking into account the broader political economic contexts. Indeed, these changes highlight the importance of asking the question of how technologies might be restructured, regulated or rearticulated, together with their socioeconomic institutions which control the labor process, in order to enable the digital to contribute to a project of emancipation through the possible abolition of scarcity and the transformation of work into an aesthetic pleasure—or even the abolition of repetitive and dangerous labor.

7   Indeed, one of the difficulties with studying software is that it requires a complete assemblage of technologies in order to work at all, what we might call its infrastructure. This might be the specific model of computer or processor that is needed to run the software, or it might be a particular operating system, or network. Here we might note that the term software hardly seems to cover the wide variety of software, hardware and historical context that needs to be included in studying what we might call the civic infrastructure[9]—but which may be similarly addressed in cognate fields like digital humanities and computational social science.

8   Indeed, the digital is in many ways the creation of a constellation of standards, canonical ways of passing around discrete information and data, that creates what we might call *witnesses* to the standard—software enforcing the disciplinary action of these standards, such as APIs (Application Programming Interfaces). Owning and controlling standards can have a political economic advantage in a post-Fordist society, and much jostling by multinational corporations and governments is exactly over the imposition of certain kinds of technical standards on the internet, or what Galloway calls protocol.[10] Indeed, "computers provide an unprecedented level of specification and control over every aspect of human society (and the rest of the environment)."[11] More specifically, the computer is a symbolic processing device that has had, and will continue to have, important repercussions for a society that increasingly depends upon knowledge and information, but it is also a historical one that can be studied comparatively and historically.

9   It is at this point we can begin to materialize the digital and ask about the specific mediations that facilitate these changes. Here, we need to be cognizant of software and digital computers connected through powerful network protocols and technologies. These infrastructural systems are generally opaque to us, and we rely on them in many cases without questioning their efficacy. Think, for example, of the number of poorly designed website forms that we are increasingly required to fill in, whether for subscriptions, job applications or college classes. These are becoming an obligatory passage point which cannot be avoided, there is no going around these computational gatekeepers, and they are the only way certain systems can even be accessed at all. They are also built of computational logics which are themselves materializations of assumptions, values and norms, often taken for granted, by the designers and programmers of the systems (e.g. related to questions of gender, race, class, etc.). We need to develop methods, metaphors, concepts and theories in relation to this software and code to enable us to think through and about these systems, as they will increasingly have important critical and political consequences. That is why being able to read behind these code-based interfaces is an important starting point to any analysis of the computational.

10  Cultural memory is now stored in computational technologies such as online photo storage, document storage etc., but also through the digitalization of culture with large-scale digital repositories of knowledge (such as the 'Newton Project' a comprehensive digital archive for Isaac Newton's papers).[12] In consequence, we are seeing a realignment of our contemporary culture. From the timeless archives of our memory institutions, to the throw-away consumer experience of disposable objects, to a softwarization of culture and the economy, we are developing new forms of memory and creativity that link together the potential for human agency and expression, and which are materialized in new technologies as a site of materialized memory and shared politics. This could be the site of a progressive politics that is linked to the importance of education and the attainment of human potential in order to develop the possibilities within each of us, and which clearly draws from the Enlightenment. This could also contribute to developing a new form of progressive post-capitalist economics with a potential for work that is creative, engaging and interesting. It could also reflect a dystopian turn, with real-time streaming systems used to build a panopticon of totally surveilled populations monitored by an all-seeing state—as the US NSA revelations showed—or citizens nudged through the application of a corporate consumerist culture that operates on the level of citizens' pre-thought.[13] This calls for a site of critique in relation to the rapid colonization by the computational and from which detailed investigations might be undertaken. This site could indeed be the university, Weizenbaum argued in 1984, that

> In mastering the programming and control of computers, we [in the university] especially could play a critical role. It may well be that no other organization is able to play this role as we are, yet no more important role may exist in science and technology today. The importance of the role stems, as has been noted, from the fact that the computer has been incorporating itself, and will surely continue to incorporate itself, into most of the functions that are fundamental to the support, protection, and development of our society.[14]

11  Indeed, the university itself, as the exemplar of learned culture and memory, and as an institution of book culture and survivor of mass media, is itself under threat from the softwarization of its underlying institutional forms and structures. Not only in terms of the rationalization of culture itself (e.g. digitalization and the resultant logics of databases and algorithms), but also through the ability, via computational systems, to construct markets and intensify principles of cost-benefit analysis to memory institutions themselves. Indeed, as Derrida commented, the university risks becoming a "branch office of conglomerates and corporations."[15] Indeed, the question remains as to whether the computer will impact the university in much the same way as Victor Hugo remarked as to the book's effect on the cathedral, *ceci tuera cela*—this will kill that.[16] Indeed, in relation to the university, Derrida writing as far back as 2000 argued,

> One of the mutations that affect the place and the nature of university work today is a certain delocalizing virtualization of the space of communication, discussion, publication, archivization... What is new, quantitatively, is the acceleration of the rhythm, the extent, and the powers of capitalization of such a virtuality... This new technical "stage" of virtualization (computerization, digitalization, virtually immediate worldwide-ization of readability, tele work, and so forth) destabilizes, as we have all experienced, the university habitat. It upsets the university's topology, disturbs everything that organizes the places defining it, namely, the territory of its fields and its disciplinary frontiers as well as its places of discussion, its field of battle, its *Kampfplatz*, its theoretical battlefield—and the communitary structure of its "campus."[17]

12  It is certainly now the case that the university will be an important battleground in the fight over the limitations or reconstructions of computational knowledges in our historical juncture, and with it the definitions of and the implications for the kinds of knowledge that were historically produced under the aegis of the arts and humanities and social science.[18] With it, questions over the required literacies to achieve majority in a computational society need to be asked in relation to these issues and something I want to explore further here. Addressing the specific problems raised by a particular literacy connected to the digital is a pressing issue. How should citizens read the digital—and to what extent citizens can and should be expected to write the digital?

13  Drawing on medium theory,[19] I want to explore the idea that to understand patterns in computational cultures we are forced to look inside the structures of digital machines— namely the notion that medial changes create epistemic shifts.[20] Further, technology and, by extension, the medium of the computal itself, create the conditions of possibility for particular cultural practices. These environments are prescribed—that is, they limit practices in certain ways that can be assessed through critical engagement with the technology. Attention to the materiality of software requires a form of reading/writing of these depths through attentiveness to the affordances of code. By attending to the ontical layers of software, that is the underlying structure and construction, we gather an insight into the substructure and machinery of software. But there is also a juridical and political economic moment here, where wealth and law are used to enforce certain control technologies, such as digital rights management (DRM), and the impossibility of opening them due to legislation that criminalizes such "readings." Software is increasingly not only mediated by its surface or interface, but through law such that the underlying mechanisms are often criminal to access. Software is therefore increasingly used/enjoyed without the encumbrance or engagement with its underlying structures due to this commodity/mechanism form—it becomes a consumption technology—enforced through law, technical means and cultural practices.

14 This has put citizens at an obvious disadvantage in a computational society that not only has historically tended to disavow the digital as a form of knowledge or practice, but also has not seen computational thinking or skills as part of the educational requirements of a well-informed citizen. Indeed, as computer power has increased, so has the tendency to emulate older media forms to provide content within simulations of traditional containers, such as "e"-books, through techniques of skeuomorphism and glossy algorithmic interface design—rather than engage specifically with the computational as such. This has enabled new computational forms to be used and accessed without the requisite computational skills to negotiate the new literary machines of computation, such as the underlying logics, structures, processes and code. In many cases today, we are unable to read what we write, and are not always the writers of the systems that are built around us. This does not seem to be the ideal conditions for the emergence of an informed and educated citizenry to engage with the challenges and dangers of a fully-softwarized society.

15 So, for example, as the old media forms, like TV, radio, film and newspapers—the traditional media of the public sphere—are digitized, there is experimentation by designers and programmers on the best form to present media content to the user. This is usually framed by the most profitable way that a passive subject position can be constructed such that its practices in relation to the interface are literally inscribed in algorithms—often conceptualized under the notion of "engagement." That is, the public sphere as a site of contestation and critique is algorithmically limited by using software that creates "delightful" interfaces that are oriented towards passivity, entertainment and the spectacular. Traditional media are softwarized in this process and the way the content used is mediated through a software interface. When transformed into software, firstly, a new media object is subject to algorithmic manipulation. One thinks here of the so-called "casual gaming" systems, that are designed to not only present a non-linear entertainment system, but also use gamification techniques to create an addictive environment for users. Game mechanics, such as badges and levels, are used to manipulate the users and serve to maximize profit for the companies using these techniques. In short, media becomes programmable. Secondly, streaming media are media built around a continuous data flow, and this will likely be the paradigmatic form of media for the future. This means that media will increasingly be subject to regular and repeating computation and the framing of the code that enables these systems will have a constitutive role in subjectivities co-constructed by them. In other words, it is the imaginary of "augmented humanity"—the notion that humanities cognitive abilities are weak and need computational reinforcement—such that the selection and comprehension of knowledge will be subject to computational pre-processing that pre-formats and aggregates before being shown to the user. In doing so computation homogenizes disparate and discontinuous knowledges into streams of data, algorithmic interfaces and dashboards.

16 This remains in the context of a society that is increasingly reliant upon a machinery that certainly does not "appear"—that is, software, algorithms, data and code. So developing our understanding of software mediation has to be understood in relation to society reaching a point at which computation is at or close to "saturation" levels. That is, that computation becomes part of the everyday life of its citizens, and as such is the norm for living within such a society. Thus we need to move to a philosophical and historical critical contextualization of computation beyond purely methodological approaches which seek to empirically map or document their research object. The kinds of ahistorical digital methods that attempt to deploy raw "facts" from which they attempt to derive "laws" from data, taking, as they do, past and present experience as though it is predictive of future experience, are increasingly useless in the diagnosis of the everyday computational present. Indeed, critical theory, as a project committed to social change, is irreconcilable with such empiricism—whether through Big Data or otherwise. When software has become crucial to the everyday operation of the society, it is clear that something important has shifted, and that we need to develop critical concepts and theories.

17  Foucault suggests that if there is a "system" or an ensemble of systems, the task is somehow to think systemic functioning outside of the perspective of the subject dominated by or in charge of the so-called system. Here we can make the link between sight and power, and of course sight itself is deployed such that the "visible" is not transparent nor hidden. Thus software and algorithms generate certain notions of truth and falsity, both in relation to knowledge itself and the very framing of the conceptual resources we deploy to think. For example, the notion of a surface interface generating a "visible" truth, and the notion of a computational, or cloud, "truth" that is delivered from the truth-machines that are mediated by the networks of power and knowledge.

18  In the first instance, a step forward can be made by exploring the processes of digital transformation of the basic categories by which a system, process, or object is understood, and especially at the early and often public moments in the "softwarization" process—before the technologies are completely formalized or "concretized." This is when, for example, an industry reconfigures and reorganizes itself in order to meet the requirements of software systems impetuous toward particular economic, structural forms and digital logics resulting in its rearticulation through the digital. I don't want to identify these technological moments as being the sole driver of economic or technical change, of course, nor the only moment of intervention, but rather highlight how these early moments in production are an important condition of mediation, for and of social labor and the economy. For example, when entities or processes are incorporated into software, they are usually transformed into *files and records*, or the "data," and *logic and algorithms*, the "software."

19  When an organization seeks to "informatize" some organizational or business logic, for example, the ways of doing and the norms associated are on the table, so to speak, as indeed are the choices in relation to how these means are encoded into algorithms, business systems and organizational logistics. This is not a trivial process and is fraught with political and economic arguments, technical challenges and breakdowns, and institutional reconfigurations and innovations. It also requires an educative dimension in relation to the framing of the uses of these systems and formats—including the harvesting of user innovation back into the system, such as shown by Twitter's absorption of "@mention names" and "hashtags" which were originally created by users themselves and now stand as crucial business logics which help to justify the market capitalization of Twitter in 2013 at $4 billion.[21]

20  The digital clearly has an instrumental dimension, in that it runs processes that are means-end oriented. But what is also radical about the digital is there is no real separation between data and execution. This epistemic aspect has many consequences in relation to the way in which data contain implicit logics, metadata and norms. For the computer and the programmer, all content and form are represented as data flows. In contrast to a factory, where one might use leather and other tools that will allow the production of commodities such as shoes, leather is not generally used to reshape the tools themselves directly. In contrast, anything structured within code and software can itself be thus transformed. So the digital is not only changing the way things are classified and the way in which things and objects are recognized by the system, but also it changes what they are and how they can be used—that is, software acts upon software ontologically. We can think of this as feedforward and feedback mechanisms that are combined with abstraction processes and layering to form an important part of software development implementation and which create rapid stages of innovation in computational systems—a process of iterative development.

21   It should hardly come as a surprise that code/software lies as a mediator between ourselves and our corporeal experiences. Software acts to disconnect the physical world from a direct coupling with our experience, mediating a looser softwarized "transmission" system of intentionality through computational interfaces. Called 'fly-by-wire' in aircraft design, in reality fly-by-wire is the condition of the computational environment we experience. This is a highly mediated existence and has been a growing feature of the (post-)digital computational world.[22] Whilst many objects remain firmly material and within our grasp, it is easy to see how a more softwarized form of augmented reality lies just beyond the horizon. Not that software isn't material, of course, certainly it is embedded in physical objects and the physical environment and requires a material carrier to function at all. Nonetheless, the materiality of software appears *uncanny* as a material and therefore more difficult to "get a grasp" of as a *material* artefact. This is partly, it has to be said, due to software's increasing tendency to hide its depths behind glass rectangular squares which yield only to certain prescribed forms of touch-based and conversational interfaces. But this is also because algorithms are always themselves doubly mediated due to their physical existence as electric pulses and flows within digital circuits which lie beyond our phenomenological experience.

22   Previously, in *The Philosophy of Software*, I outlined the emergence of computationality as an *ontotheology* drawing on the work of Heidegger.[23] I argued that computationality is a specific historical epoch defined by a certain set of computational knowledges, practices, methods and categories. Computationality which reads through Heideggerian categories can be understood as creating a new ontological "epoch" or a new historical constellation of intelligibility. With the notion of ontotheology, Heidegger is following Kant's argument that intelligibility is a process of filtering and organizing a complex overwhelming world by the use of "categories," Kant's "discursivity thesis." Heidegger historicizes Kant's cognitive categories arguing that there is a "succession of changing historical ontotheologies that make up the 'core' of the metaphysical tradition. These ontotheologies establish 'the truth concerning entities as such and as a whole,' in other words, they tell us both what and how entities are—establishing both their essence and their existence."[24] Metaphysics, grasped ontotheologically, "temporarily secures the intelligible order' by understanding it 'ontologically,' from the inside out, and 'theologically' from the outside in, which allows the formation of an epoch, a 'historical constellation of intelligibility which is unified around its ontotheological understanding of the being of entities."[25]

23  Thus, as an ontotheology, computationality is a central, effective, increasingly dominant system of meanings and values that become operative and which is not merely abstract but which is organized and lived. Thus computationality cannot be understood at the level of mere opinion or manipulation—it is not merely ideological in form. It is related to a whole body of computational practices and expectations, for example the assignment of energy towards particular projects, the ordinary understanding of the 'nature' of humans, and of the world. This set of meanings and values are experienced as practices which appear as reciprocally confirming, repeated and predictable and also used to describe and understand the world—in some cases, software even becomes an explanatory form of explanation itself.[26] This analysis also draws from previous theoretical work undertaken by Horkheimer and Adorno, particularly in relation to the way in which the domination of nature is entangled with the "mastery over human nature, the repression of impulse, but also the mastery over other humans."[27]

24  We experience algorithms in their performances through practices that rely on computers, but also on screenic representation and so forth. Code/software and the processes and agency they engender, are the paradigmatic cases of computationality. Indeed, they present us with a set of research entities (code-objects) which are located at all major junctures of modern society and are unique in enabling modern society but also raising the possibility of reading and understanding the present situation of computationality, as a massive distributed network of computation which penetrates society at all levels. But additionally the computal operates in a more essential sense, structuring categories, classifications and so forth, which "leak" out of computational systems and become absorbed into cultural and institutional practices, shared encounters, memories, norms and values.

25  Additionally, any study of computer code has to acknowledge that the performativity of software is in some way linked to its location in a capitalist economy. Code costs money and labor to produce and once it is written requires continual inputs of energy, maintenance and labor to keep functioning. Thus code is socially constructed, historically specific and more or less socially embedded in broader networks of social relations and institutional ensembles.[28] It is crucial that the materiality and ownership of code be understood and the constraints that operate upon the production, distribution and consumption of code as software be noted. This has important implications when it is understood that much of the code that supports the Internet, even though it is free software or open source, actually runs on private computer systems and networks.[29] Understanding the theoretical, empirical and political economic aspects of 'computational cultures' in relation to the so-called knowledge economy, particularly through the lens of critical theory, requires us to engage with this computational dimension of the digital. Further, computation is the logic of the "creative" economy and to understand the cultural outputs of computational structures (sometimes referred to as the "softwarization of culture") we need a critical theory that can contribute to the understanding of the computational.

26  This applies also to the notion of not only aggregating objects and human beings as networks using software, but also treating human beings as components or objects *of* a computational system. Indeed, this is indicative of the kind of thinking that is prevalent in computational design. Production or consumption are treated by the creation of code-objects to represent activities in everyday life and translate them internally into a form the computer can understand. In many ways this is a discretization of human activity, but it is also the dehumanization of people through a computation layer used to mediate the use of social labor more generally. This also demonstrates how the user is configured through code-objects as producer, consumer, worker or audience, a new kind of multiple subject-position that is disciplined through computational interfaces and algorithmic control technologies. But it also serves to show how the interface reifies the social labor undertaken behind the surface, such that the machinery may be literally millions of humans "computing" the needs to the software, all without the end-user being aware of it. In this case it is not that the machinery represents what Marx called "dead labor," but in fact that it mediates living labor invisibly into the machinery of computation. Indeed, this is an example of where continuous computation serves to hide social labor such that workers are hidden "behind web forms and APIs [which] helps employers see themselves as builders of innovative technologies, rather than employers unconcerned with working conditions."[30]

27  These computational systems therefore enable the assemblage of new social ontologies and the corresponding social epistemologies and logistics that we increasingly take for granted in computational society, for example in Wikipedia, Facebook, and Twitter. The extent to which computational devices, and the computational principles on which they are based and from which they draw their power, have permeated the way we use and understand knowledges in everyday life is remarkable, had we not already discounted and backgrounded their importance.

28  In the case of computational ontologies, and the use of computational concepts more widely within our ontological and everyday understanding of life, the question is: to what extent do these computational categories perform not merely as what Adorno called "wretched" cover-concepts? Indeed, do they have the possibility of generatively making possible contradictions that facilitate critical thought, within what we are calling here computationality, as emphatic conceptual resources? To look more closely at the computal and computational ontologies it helps to think through the distinction introduced by Adorno between what he called "cover-concepts" and their distinction from "emphatic concepts."[31] That is,

> A cover-concept is one which can be used to limit the members of a set. It is descriptive. But an emphatic concept is one which has inside it a promise. It is a promise which cannot be cut out of the concept without changing it. So that the concept of "art," it could be suggested, is not merely a cover-concept. It does not signify a certain set of properties, any object possessing which could count as an instance of the concept. To call something art is always not only to describe something but also to evaluate it.[32]

29  Adorno argues that emphatically conceived, a concept, is "one that is not simply the characteristic unit of the individual object from which it was abstracted."[33] That is, like the concept of freedom, these emphatic concepts are not merely descriptive, and therefore "arbitrarily diminished," instead there is a "more" of the concept, as it were, which offers the possibility of generating a contradiction between the concept of freedom and its realization, and therefore the possibility of critical thought itself. Concepts such as "freedom, humanity, and justice are what Adorno calls 'emphatic' concepts in the sense that they are ineliminably both prescriptive and descriptive."[34]

30  This is something that I have been thinking about too in relation to the emphatic concepts of education and digital *Bildung*.[35] I would like to suggest that *iteracy* might serve as a signifier for the range of skills used for understanding computation—as indeed literacy (understanding texts) and numeracy (understanding numbers) do in a similar context. That is, iteracy is specifically the practice or being able to read and write digital texts and computational processes, and contained underneath the more essential notion of digital *Bildung*.[36] Here, digital *Bildung* is understood as the totality of education in the university of the digital age, not as a subject trained in a vocational fashion to perform instrumental labor, nor as a subject skilled in a national literary culture, but rather as a subject that can reconcile the information that society is now producing at increasing rates, and who understands new methods and practices of critical reading (such as code, data visualization, patterns, narrative) and taught using new and old methods of pedagogy to facilitate it.[37]

31  So digital *Bildung* would include the practices of iteracy and would build on them to facilitate a broader humanistic or critical education. Here, iteracy is defined broadly as communicative competence in reading, writing and executing computer code. This calls for a different kind of relationship in the creation and dissemination of knowledge in the university, perhaps a reinvigorated form of educational research and teaching which is opposed to the depressingly service-oriented vocationalism and mass-delivery platforms that have dominated much discussion of university imaginaries. When we think about the changes wrought by the digital technologies that are increasingly structuring our lives, it is important to remember the warnings that Joseph Weizenbaum gave for the university:

> The function of the university cannot be to simply offer prospective students a catalog of "skills" from which to choose… Surely the university should look upon each of its citizens, students and faculty alike, first of all as human beings in search of—what else to call it?—truth, and hence in search of themselves.[38]

32  Having a grasp of the basic principles of iteracy as a critical orientation towards the computational is crucial for reading code and for undertaking a critical approach in the digital age. This is because the ubiquity of computation, and the way in which norms and values are delegated into algorithms creates an invisible site of power, which also has agentic power. It is also the case that part of the critique of software has to be the ability to unbuild these systems, to take them apart and to provide critical "readings" of them. We live in deeply computational societies with ways of working with software that calls for new cognitive maps. With the increase in ubiquity of these computer systems in all aspects of life, it is likewise important that citizens have the skills to understand and critique them.

33  Clearly, we have to be careful not to narrow iteracy to only formal programming knowledge. Indeed, I have found it very useful to explain to students that they are 'programming' a computer when they set an alarm on their iPhone or negotiate a menuing system in Photoshop. This highlights that when using/programming a computer it is *black-boxes all the way down*—and that this layering within computational technologies is part of computational structures writ large—but also that we need to be able to potentially open these black-boxes all the way down.[39] Increasingly, I think "iteracy" will be as crucial for operating in this computational culture—especially considering the ontologies that are delegated into the devices that surround us take for granted certain computational principles of operation, such as real-time data and media streams—as numeracy and literacy have been.

34  Iteracy, therefore, also refers to the ability to critically read, write and understand processes, that is, following Wardrip-Fruin's notion of "process descriptions."[40] So there are, perhaps, two levels of writing taking place here, the textual and the processual. This highlights the way in which we can think of this as a depth model of computation as digital writing, (1) code/text/data (deep) and (2) the process/screenic (flat). This is a simplification, however it is a useful heuristic for thinking about the kinds of things we need to take account of in teaching and researching computational media. This also helps draw attention both to reading code and towards reading processes.[41]

35  Indeed, something akin to the hermeneutic circle is needed here, whereby the code is understood not merely through a close reading of the text, but by running it, observing its operation and the processes it institutes, introducing breakpoints and "print to screen" functions to see inside the code whilst it is running, such as through the use of tests. Programmers, who have iteracy by education and habit, are able to jump between these perspectives on the code (code as text, code as process, code as whole system), seamlessly backwards and forwards as they develop knowledge and understanding of the code. This is similar to a notion of a "fusion of horizons" but needs to be supplemented by critical readings that explore how code-objects exist in a historical, political and socioeconomic context and usually with a certain aim or intention (whether achieved or not).

36  In particular, I want to relate this to the notion of a holistic digital education, or digital *Bildung* for the university. More specifically as methods and approaches related to critical inquiry of the computal.[42] I do think that iteracy has some heuristic advantages over terms like 'code literacy', 'digital literacy', 'information literacy', and so forth, especially the connotations that iteracy has with iteration, a key part of how code functions are read and written. Some of the components of such an approach could include: (i) *critical computational thinking,* or being able to devise and understand the way in which computational systems work to be able to reflexively read and write the code associated with them. For example abstraction, pipelining, hashing, sorting, etc.[43]. (ii) *understanding algorithms*: specifically algorithmic nature of computational work, e.g. recursion, iteration, discretization, etc. (iii), understanding the significance and importance of *data and models* particularly of data, information and knowledge and their relationships to models in computational thinking. (iv) critical technical practices in *reading and writing code* which require new skills to enable the reader/programmer to make sense of and develop code in terms of modularity, data, encapsulation, naming, commentary, loops, recursion, etc. (v) *learning programming languages* as understanding one or more concrete programming languages enables the student to develop a comparative approach and hones the skills associated with iteracy, for example, procedural, functional, object-oriented languages, etc. (vi) developing skills related to appreciating *code aesthetics*, that is the aesthetic dimension of code, software and algorithms, including notions of 'beautiful code' and 'elegance' as key concepts,[44] but also the question of the digital and aesthetics in relation to new media art and new digital aesthetics.[45]

37  Thus the university has to engage not just with the traditional knowledges that it has become accustomed to, and institutionalized within its disciplinary structure, but also with computational knowledge more broadly. In many cases computation has become too important as a framework of understanding society, and as a condition of possibility for political and social engagement, to be left outside of the humanities and social sciences. The call for a digital *Bildung* is for computation to be part of the critical traditions of the arts and humanities, the social sciences and the university as a whole. Whether this will best be achieved through a disciplinary formation, such as critical digital humanities, or through a more trans-disciplinary program of multiplicity throughout the arts and humanities and social sciences remains to be seen. It is clear, however, that increasingly computational knowledges are becoming traditional knowledges both in terms of the articulation and mediation of culture and its archives, but also as the means of reading and understanding them, both now and in our increasingly computational futures.

## BIBLIOGRAPHY

ADORNO, Theodor W., *Negative Dialectics*, London/New York Routledge, 2004 [1966].

BERRY, David M., *Copy, Rip, Burn: The Politics of Copyleft and Open Source*, London, Pluto Press, 2008.

BERRY, David M., *The Philosophy of Software: Code and Mediation in the Digital Age*, London, Palgrave Macmillan, 2011.

BERRY, David M., "The social epistemologies of software," *Social Epistemology*, 26(3-4), 2012: 379–398.

BERRY, David M., *Critical Theory and the Digital*, New York, Bloomsbury Academic, 2014.

BERRY, David M. (ed.), *Understanding Digital Humanities*, London, Palgrave Macmillan, 2012.

BERRY, David M. and DIETER, Michael (eds.), *Postdigital Aesthetics: Art, Computation and Design*, Basingstoke, Palgrave Macmillan, 2015.

CHUN, Wendy Hui Kyong, "On software, or the persistence of visual knowledge," *Grey Room*, 18, 2004: 26–51.

CHUN, Wendy Hui Kyong, *Programmed Visions: Software and Memory*, Cambridge (MA), MIT Press, 2011.

DAMES, Nicholas, "This will kill that" [on line], *n+1*, August 11, 2010, available at <http://nplusonemag.com/this-will-kill-that> [accessed 12/28/2013].

DERRIDA, Jacques, *Without Alibi*, P. Kamuf (ed., intr., transl.), Stanford, Stanford University Press, 2000.

DREYFUS, Hubert L. and RABINOW, Paul, *Michel Foucault: Beyond Structuralism and Hermeneutics*, Chicago, University of Chicago Press, 1982.

ERNST, Wolfgang, "Media archaeography: method and machine versus the history and narrative of media," *in* Id., *Digital Memory and the Archive*, J. Parikka (ed.), Minneapolis, University of Minnesota Press, 2013: 55–73.

FARRELL, Maureen, "Twitter's selloff accelerates: market cap falls by $4 billion" [on line], *The Wall Street Journal*, December 27, 2013, available at <http://blogs.wsj.com/moneybeat/2013/12/27/twitters-selloff-accelerates-market-cap-falls-by-4-billion/> [accessed 12/30/2013].

FULLER, Matthew, "Introduction," *in* Id., *Software Studies: A Lexicon*, Cambridge (MA), MIT Press, 2008: 1–14.

GALLOWAY, Alexander R., *Protocol: How Control Exists After Decentralization*, Cambridge (MA), MIT Press, 2006.

GALLOWAY, Alexander R., "Language wants to be overlooked: on software and ideology," *Journal of Visual Culture*, 5(3), 2006: 315–331.

GOLUMBIA, David, *The Cultural Logic of Computation*, Cambridge (MA), Harvard University Press, 2009.

HEIDEGGER, Martin, *Überlieferte Sprache und Technische Sprache*, H. Heidegger (ed.), St. Gallen, Erker, 1989 [1962] [english version: "Traditional Language and Technological Language," W. Torres Gregory (trans.), *Journal of Philosophical Research*, 23, 1998: 129–145].

IRANI, Lilly C. and SILBERMAN, M. Six, "Turkopticon: interrupting worker invisibility in Amazon mechanical Turk" (CHI 2013 Proceedings of the SIGCHI Conference on Human Factors in Computing Systems, Paris, April 27–May 2, 2013), New York, ACM, 2013: 611–620, available at <http://wtf.tw/text/turkopticon.pdf> [accessed 07/10/2013].

JARVIS, Simon, *Adorno: A Critical Introduction*, Cambridge, Polity Press, 1998.

JARVIS, Simon, "The truth in verse? Adorno, wordsworth, prosody," *in* D. Cunningham and N. Mapp (eds.), *Adorno and Literature*, New York, Continuum, 2006: 84–98.

KITTLER, Friedrich A., *Literature, Media, Information Systems: Essays*, J. Johnston (ed., intr.), Amsterdam, OPA, 1997.

MANOVICH, Lev, "Media After Software," *Journal of Visual Culture*, 12(1), 2013: 30–37.

MANOVICH, Lev, "Introduction," in *Software Takes Command: Extending the Language of New Media*, New York/London, Bloomsbury, 2013: 1–43.

MITCHAM, Carl, "The importance of philosophy to engineering," *Teorema*, XVII(3), 1998: 27–47.

ORAM, Andy and WILSON, Greg, *Beautiful Code: Leading Programmers Explain How They Think*, London, O'Reilly Media.

SCHECTER, Darrow, *The History of the Left from Marx to the Present: Theoretical Perspectives*, New York, Continuum, 2007.

SIEGERT, Bernhard, "Cultural techniques: or the end of the intellectual postwar era in German media theory," *Theory, Culture & Society*, 30(6), 2013: 48–65.

STAR, Susan Leigh, "The ethnography of infrastructure," *American Behavioral Scientist*, 43(3), 1999: 377–391.

THOMSON, Iain, "Understanding technology ontotheologically, or: the danger and the promise of Heidegger, an American perspective," *in* J.K.B Olsen, E. Selinger and S. Riis (eds.), *New Waves in Philosophy of Technology*, London, Palgrave Macmillan, 2009: 146–166.

WARDRIP-FRUIN, Noah, *Expressive Processing: Digital Fictions, Computer Games, and Software Studies*, Cambridge (MA), MIT Press, 2009.

WEIZENBAUM, Joseph, *Computer Power and Human Reason: From Judgement to Calculation*, London, Penguin Books, 1984 [1976].

WING, Jeannette M., "Research notebook: computational thinking—what and why?" [on line], *thelink.*, March 6, 2011, available at <http://link.cs.cmu.edu/article.php?a=600> [accessed 09/16/2011].

## NOTES

**1.** David M. Berry, *The Philosophy of Software: Code and Mediation in the Digital Age*, London, Palgrave Macmillan, 2011; David M. Berry, *Critical Theory and the Digital*, New York, Bloomsbury Academic, 2014.

**2.** By *computal* I refer to computational techniques and practices, digital media, code, algorithms and software more generally.

**3.** David M. Berry (ed.), *Understanding Digital Humanities*, London, Palgrave Macmillan, 2012.

**4.** David M. Berry, "The social epistemologies of software," *Social Epistemology*, 26(3-4), 2012: 379–398.

**5.** Carl Mitcham, "The importance of philosophy to engineering," *Teorema*, XVII(3), 1998: 43.

**6.** For a discussion, see David M. Berry, *Copy, Rip, Burn: The Politics of Copyleft and Open Source*, London, Pluto Press, 2008; *Id.*, *The Philosophy of Software...*, *op. cit.*

**7.** For example, see the work being undertaken in Software Studies: David M. Berry, *The Philosophy of Software...*, *op. cit.*; Wendy Hui Kyong Chun, "On software, or the persistence of visual knowledge," *Grey Room*, 18, 2004: 26–51; *Id.*, *Programmed Visions: Software and Memory*, Cambridge (MA), MIT Press, 2011; Matthew Fuller, "Introduction," *in* Id., *Software Studies: A Lexicon*, Cambridge (MA), MIT Press, 2008: 1–14; Alexander R. Galloway, "Language wants to be overlooked: on software and ideology," *Journal of Visual Culture*, 5 (3), 2006: 315–331; Lev Manovich, "Media After Software," *Journal of Visual Culture*, 12(1), 2013: 30–37; *Id.*, "Introduction," in *Software Takes Command: Extending the Language of New Media*, New York/London, Bloomsbury, 2013: 1–43.

**8.** Hubert L. Dreyfus and Paul Rabinow, *Michel Foucault: Beyond Structuralism and Hermeneutics*, Chicago, University of Chicago Press, 1982: 216.

**9.** See Susan Leigh Star, "The ethnography of infrastructure," *American Behavioral Scientist*, 43(3), 1999: 377–391.

**10.** Alexander R. Galloway, *Protocol: How Control Exists After Decentralization*, Cambridge (MA), MIT Press, 2006.

**11.** David Golumbia, *The Cultural Logic of Computation*, Cambridge (MA), Harvard University Press, 2009: 216.

**12.** See <http://www.newtonproject.sussex.ac.uk/>.

**13.** The relationship between computation, surveillance and critical cryptographic practices is something I discuss in more detail in *Critical Theory and the Digital* (New York, Bloomsbury Academic, 2014).

**14.** Joseph Weizenbaum, *Computer Power and Human Reason: From Judgement to Calculation*, London, Penguin Books, 1984 [1976]: 242.

**15.** Jacques Derrida, *Without Alibi*, P. Kamuf (ed., intr., transl.), Stanford, Stanford University Press, 2000: 241.

**16.** "*Ceci tuera cela*": the famous slogan of Claude Frollo, the archdeacon of Notre-Dame in Victor Hugo's *Notre-Dame de Paris*, as he touches a printed book and glances nostalgically at the cathedral towers. "This will kill that" (Nicholas Dames, "This will kill that" [on line], *n+1*, August 11, 2010, available at <http://nplusonemag.com/this-will-kill-that> [accessed 12/28/2013]).

**17.** Jacques Derrida, *Without Alibi*, op. cit.: 210.

**18.** David M. Berry (ed.), *Understanding Digital Humanities*, op. cit.

**19.** Wolfgang Ernst, "Media archaeography: method and machine versus the history and narrative of media," *in* Id., *Digital Memory and the Archive*, J. Parikka (ed.), Minneapolis, University of Minnesota Press, 2013: 55–73; Friedrich A. Kittler, *Literature, Media, Information Systems: Essays*, J. Johnston (ed., intr.), Amsterdam, OPA, 1997; Bernhard Siegert, "Cultural techniques: or the end of the intellectual postwar era in German media theory," *Theory, Culture & Society*, 30(6), 2013: 48–65.

**20.** See David M. Berry, *The Philosophy of Software…*, op. cit.

**21.** Maureen Farrell, "Twitter's selloff accelerates: market cap falls by $4 billion" [on line], *The Wall Street Journal*, December 27, 2013, available at <http://blogs.wsj.com/moneybeat/2013/12/27/twitters-selloff-accelerates-market-cap-falls-by-4-billion/> [accessed 12/30/2013].

**22.** David M. Berry and Michael Dieter (eds.), *Postdigital Aesthetics: Art, Computation and Design*, Basingstoke, Palgrave Macmillan, 2015.

**23.** David M. Berry, *The Philosophy of Software…*, op. cit.

**24.** Iain Thomson, "Understanding technology ontotheologically, or: the danger and the promise of Heidegger, an American perspective," *in* J.K B Olsen, E. Selinger and S. Riis (eds.), *New Waves in Philosophy of Technology*, London, Palgrave Macmillan, 2009: 149–150.

**25.** *Ibid.*: 150.

**26.** Wendy Hui Kyong Chun, *Programmed Visions…*, op. cit.

**27.** Darrow Schecter, *The History of the Left from Marx to the Present: Theoretical Perspectives*, New York, Continuum, 2007: 27.

**28.** I have explored some of these issues in *Copy, Rip, Burn: The Politics of Copyleft and Open Source* (London, Pluto Press, 2008) and *The Philosophy of Software: Code and Mediation in the Digital Age* (London, Palgrave Macmillan, 2011).

**29.** David M. Berry, *Copy, Rip, Burn…*, op. cit.

**30.** Lilly C. Irani and M. Six Silberman, "Turkopticon: interrupting worker invisibility in Amazon mechanical Turk" (CHI 2013 Proceedings of the SIGCHI Conference on Human Factors in Computing Systems, Paris, April 27–May 2, 2013), New York, ACM, 2013: 611–620, available at <http://wtf.tw/text/turkopticon.pdf> [accessed 07/10/2013].

**31.** Theodor W. Adorno, *Negative Dialectics*, London/New York Routledge, 2004 [1966]: 148-151.

**32.** Simon Jarvis, "The truth in verse? Adorno, wordsworth, prosody," *in* D. Cunningham and N. Mapp (eds.), *Adorno and Literature*, New York, Continuum, 2006: 88.

**33.** Theodor W. Adorno, *Negative Dialectics*, op. cit.: 150.

**34.** Simon Jarvis, *Adorno: A Critical Introduction*, Cambridge, Polity Press, 1998: 66. Some of the more obvious emphatic concepts associated with technology include: progress, freedom, information, communication, distribution, open, free, education, meritocracy, democracy, liberty, rationality, intelligence, etc. This is not an exhaustive list, but demonstrates that computation draws on emphatic as well as cover concepts and the potential for transcending instrumentalism through the emphatic is a latent possibility within technological orders of discourse and materialities as much as in other spheres.

**35.** David M. Berry, *The Philosophy of Software…, op. cit.*

**36.** *Ibid.*: 20-26.

**37.** *Ibid.*: 168. Currently I teach a module called Theory and Practice of Interactive Media at the University of Sussex that attempts to bring many of these ideas together by closely relating the question of opening the black box of computation with the issue of practices related to reading and writing the digital. Theoretically informed by software studies, media archaeology, theories of aesthetics, digital humanities and political activism the module engages with students' need to think critically in and against the digital.

**38.** Joseph Weizenbaum, *Computer Power and Human Reason…, op. cit.*: 278.

**39.** See David M. Berry, *Critical Theory and the Digital*, *op. cit.*, for a discussion of how the laminated structure of computation help us develop a language for understanding and describing our object(s) of study at an 'appropriate' ontological level, such as the Physical, Logical, Codal, Interactional, Logistical, and Individuational.

**40.** Noah Wardrip-Fruin, *Expressive Processing: Digital Fictions, Computer Games, and Software Studies*, Cambridge (MA), MIT Press, 2009.

**41.** Cf. David M. Berry, *Critical Theory and the Digital*, *op. cit.*: 58.

**42.** David M. Berry, *Critical Theory and the Digital*, *op. cit.*

**43.** See Jeannette M. Wing, "Research notebook: computational thinking—what and why?" [on line], *thelink.*, March 6, 2011, available at <http://link.cs.cmu.edu/article.php?a=600> [accessed 09/16/2011].

**44.** See Andy Oram and Greg Wilson, *Beautiful Code: Leading Programmers Explain How They Think*, London, O'Reilly Media; David M. Berry, *Critical Theory and the Digital*, *op. cit.*

**45.** David M. Berry and Michael Dieter (eds.), *Postdigital Aesthetics…, op. cit.*

## ENDNOTES

**i.** The title of this chapter owes a debt to Martin Heidegger, *Überlieferte Sprache und Technische Sprache*, H. Heidegger (ed.), St. Gallen, Erker, 1989 [1962] [english version: "Traditional Language and Technological Language," W. Torres Gregory (trans.), *Journal of Philosophical Research*, 23, 1998: 129–145].

## ABSTRACTS

If critical approaches are to remain relevant in a computational age, then philosophy must work to critique and understand how the materiality of the modern world is normatively structured using computation and the attendant imaginaries made possible for the reproduction and transformation of society, economy, culture and consciousness. This call is something we need to respond to in relation to the contemporary reliance on computational forms of knowledge and practices and the co-constitution of new computational subjectivities. This chapter argues that to comprehend the digital we must, therefore, know it from the inside, we must know its formative processes. We must materialize the digital and ask about the specific mediations that are made possible in and through computation, and the infrastructural systems which are built from it. This calls for computation and computational thinking to be part of the critical traditions of the arts and humanities, the social sciences and the university as a whole, requiring new pedagogical models that are able to develop new critical faculties in relation to the digital.

Pour que les approches critiques restent pertinentes à l'ère computationelle, la philosophie doit développer une compréhension critique des structures normatives du monde moderne, à travers la manière dont les capacités de calcul et leurs imaginaires associés rendent possible la reproduction et la transformation de la société, de l'économie, de la culture et de la conscience. Cet appel est une question à laquelle nous devons répondre par rapport à la dépendance contemporaine à l'égard des formes informatiques de la connaissance et des pratiques et à la co-constitution de nouvelles subjectivités informatiques. Ce chapitre soutient que pour comprendre le numérique, nous devons le connaître de l'intérieur, à travers ces processus de formations. Nous devons matérialiser le numérique et poser des questions sur les médiations spécifiques qui sont rendues possibles dans et par le calcul, et les infrastructures qui sont construits à partir de celui-ci. Cela implique que le calcul et la pensée computationnelle fassent partie des traditions critiques des arts et des sciences humaines, des sciences sociales et de l'université dans son ensemble, nécessitant de nouveaux modèles pédagogiques capables de développer de nouvelles facultés critiques par rapport au numérique.

## INDEX

## AUTHOR

DAVID M. BERRY

**David M. Berry** is Professor of Digital Humanities and co-Director of the Sussex Humanities Lab. His recent books include *Critical Theory and the Digital* (2014), *Postdigital Aesthetics: Arts, Computation and Design* (2015) and *Digital Humanities Knowledge and Critique in a Digital Age* (2017, with Anders Fagerjord). He was recently awarded a British Academy Mid-Career Fellowship for his new research on "Reassembling the University: The Idea of a University in a Digital Age."